\documentclass[showpacs,groupedaddress,twocolumn]{revtex4}
\usepackage{CJK}

\usepackage{graphicx}  
\usepackage{amsfonts,amssymb,amsmath,amsthm}
\usepackage{bbold,mathtools}
\usepackage{verbatim} 
\usepackage{hyperref} 

	\def\PRLshow#1{}
	\def\paper{paper}

	\newtheorem{theorem}{Theorem}
	\newtheorem{lemma}[theorem]{Lemma}

\theoremstyle{definition}

\theoremstyle{remark}

\newcommand{\bi}[2]{\binom{#1}{#2}}
\DeclareMathOperator{\tr}{Tr}

\DeclareMathOperator{\wt}{wt}

\DeclareMathOperator{\supp}{supp}

\def\>{\rangle}

\def\<{\langle}

\def\bx{{\bf x}}

\def\d{\delta}
\def\eps{\epsilon}

\def\bbC{\mathbb{C}}

\def\cA{\mathcal{A}}
\def\cB{\mathcal{B}}
\def\cC{\mathcal{C}}

\def\cI{\mathcal{I}}

\def\cR{\mathcal{R}}

\def\fD{\mathfrak{D}}

\def\fK{\mathfrak{K}}

\def\tA{{\bf{A}}}
\def\tB{{\bf{B}}}

\def\tD{{\bf{D}}}
\def\tE{{\bf{E}}}
\def\tF{{\bf{F}}}
\def\tG{{\bf{G}}}

\def\tI{{\bf{I}}}

\def\tK{{\bf{K}}}

\def\tP{{\bf{P}}}

\def\tR{{\bf{R}}}

\def\tU{{\bf{U}}}
\def\tV{{\bf{V}}}

\def\tX{{\bf{X}}}
\def\tY{{\bf{Y}}}
\def\tZ{{\bf{Z}}}

\def\b{{\beta}}
\def\d{\delta}
\def\eps{\epsilon}

\begin{document} 

\title{Permutation-invariant quantum codes}%
\author{Yingkai \surname{Ouyang}  }
\affiliation{University of Waterloo, Waterloo, Ontario, Canada}
\affiliation{Singapore University of Technology and Design, Singapore}
\email{yingkai\_ouyang@sutd.edu.sg}

\begin{abstract}
A quantum code is a subspace of a Hilbert space of a physical system chosen to be correctable against a given class of errors, where information can be encoded. Ideally, the quantum code lies within the ground space of the physical system. When the physical model is the Heisenberg ferromagnet in the absence of an external magnetic field, the corresponding ground-space contains all permutation-invariant states. We use techniques from combinatorics and operator theory to construct families of permutation-invariant quantum codes. These codes have length proportional to $t^2$; one family of codes perfectly corrects arbitrary weight $t$ errors, while the other family of codes approximately correct $t$ spontaneous decay errors.
The analysis of our codes' performance with respect to spontaneous decay errors utilizes elementary matrix analysis, where we revisit and extend the quantum error correction criterion of Knill and Laflamme, and Leung, Chuang, Nielsen and Yamamoto.
\end{abstract}

\maketitle

\section{Introduction}
A quantum bit (qubit) is a fundamental resource required in many quantum information theoretic tasks, such as in quantum cryptographic protocols \cite{BB84} and in quantum computers \cite{nielsen-chuang}.
To combat decoherence, a two-level system (qubit) may be encoded as a quantum code, a subspace of the Hilbert space of a physical system.
 Ideally, the quantum code lies within the ground space of the physical system. A well studied example of such families of quantum codes are Kitaev's toric code and surface codes \cite{Kit03,KiB02}, where the underlying Hamiltonian of the physical system has 4-particle interactions or requires the use of Majorana fermions.
Kitaev's physical model \cite{Kit03} would be easy to implement, if not for the difficulties in realizing Majorana fermions \cite{Kit01,KiB02,Kit03} or four-way interactions in nature.
One might then wonder, if simple pairwise interactions can be used directly to design quantum codes.
Indeed, many such models have been studied extensively in the context of topological codes \cite{Kit06,Bom10,BKM09,BFBD11,OCB11}, and in this paper, we pay special attention to the ferromagnetic Heisenberg model.

The exchange interaction, arising from the inherent indistinguishability of identical particles and mainly Coulomb interactions \cite{Dirac1929,Blundell}, is a naturally abundant pairwise interaction. 
Heisenberg models \cite{Heisenberg1928,Blundell} describe physical systems with dynamics dominated by exchange interactions, such as many electron systems.
In the absence of external magnetic fields, Heisenberg models have Hamiltonians of the form
\[
H = -2 \sum_{\substack{ e= \{i,j\}\\ i < j}} J_{e} {\bf S}_i \cdot {\bf S}_j
= -  \sum_{ e  } J_{e} (\pi_{e} - \frac 1 2 \mathbb 1), 
\]
where $\mathbb 1$ is the identity operator, the indices $i$ and $j$ label the particles in the system, and $e=\{i,j\}$ labels the exchange interactions in the system.
Here $J_e$ and ${\bf S}_i$ denote the exchange constants and the vector of spin operators respectively. 
Since exchange operators essentially swap particles (see ex.~1.9 of Ref.~\cite{Blundell} or Ref.~\cite{Dirac1929}), the Heisenberg Hamiltonian $H$ can be expressed in terms of
the swap operators $\pi_e$ that swap the spin-$\frac 1 2$ particles $i$ and $j$.
We consider ferromagnetic Heisenberg models (all non-zero exchange constants are positive) of spin-$\frac{1}{2}$ particles, 
where every pair of particles interacts at least indirectly via a connected chain of interactions.

The ground space of Heisenberg ferromagnets necessarily contains the space of all permutation-invariant states.
To see this, note that any permutation-invariant state $|\psi\>$ is invariant under swap, in the sense that for all interactions $e$,  
$\pi_e |\psi\> = |\psi\>$. 
Let $J =   \sum_e J_e$, so that $H= -\sum_e J_e \pi_e +  \frac J 2 \mathbb 1.$
Then we have
\begin{align}
(H - \frac J 2 \mathbb 1) |\psi\> = - J |\psi\>. \notag
\end{align}
The non-negativity of the exchange constants $J_e$ implies that $J$ is an upper bound on the spectral norm of $H - \frac J 2 \mathbb 1$, and it follows from the above eigenvalue equation that $-J$ is the smallest eigenvalue of $H- \frac J 2 \mathbb 1$. Hence any permutation-invariant state $|\psi\>$ is a ground-state of $H$.
This motivates our study of permutation-invariant (PI) codes, since such codes are necessarily in the ground space of any Heisenberg ferromagnet. 

Previously Ruskai and Pollatsek studied several \cite{Rus00,PoR04} PI codes using the Knill-Laflamme error correction conditions \cite{KnL97}. 
Of special note is Ruskai's {9-qubit} PI code that corrects exactly one arbitrary error \cite{Rus00}, which is also precisely the completely symmetrized {9-qubit} Shor code \cite{ShorCode}. 
In this paper, we prove that the completely symmetrized extensions of the Shor code, and the infinite family of the completely symmetrized versions of the Bacon-Shor codes \cite{Bac06} of length $(2t+1)^2 \sim 4t^2$, are PI codes that correct $t$-qubit errors for all positive integers $t$. 
We also prove that a length $(t+1)(3t+1)+t \sim 3t^2$ PI code suffices to correct $t$ spontaneous decays. 

For any positive integers $g,m$ and $n$ with $m \ge gn$, 
our quantum code encodes a qubit in 
$m$ particles with logical basis states 
\begin{align}
 |\pm_L\>  &:=  \sum_{\ell = 0}^{n} \frac{(\pm 1)^\ell}{\sqrt{2^{n}}}
	 \sqrt{\bi{n}{\ell}}  |D^m_{g \ell } \> \label{eq:PI-code}.
\end{align}
In the notation of \cite{BGu13,MHT12,TGG09}, $|D^m_w\>$ is a {\em Dicke state}, which is a normalized PI state on $m$ qubits with a single excitation on $w$ qubits. We say that such a Dicke state has {\em weight} $w$.
On spin-half particles, the Dicke state $|D^m_w\>$ may also be interpreted as the uniform superposition of all states with exactly $w$ particles in the spin-up configuration, and $m-w$ particles in the spin-down configuration. 
For example the Dicke state $|D^4_2\>$ can be written as 
\begin{align}
&\frac{|0011\>  + |0110\>+ |1100\> + |1001\> + |1010\> + |0101\>}{\sqrt{6}} \notag ,\\
=& \frac{ 
|\!\downarrow \downarrow \uparrow \uparrow \> 
+|\!\downarrow \uparrow \uparrow \downarrow \> 
+|\!\uparrow \uparrow \downarrow \downarrow \> 
+|\!\uparrow \downarrow \downarrow \uparrow \> 
+|\!\uparrow \downarrow  \uparrow \downarrow \> 
+|\!\downarrow \uparrow \uparrow \downarrow \> 
}{\sqrt{6}} .\notag
\end{align}
Here the {\em code gap} $g$ and the {\em code occupancy} $n$ are positive integers; 
our quantum code lies within the span of $(n+1)$ Dicke states with weights that are consecutive multiples of $g$ apart starting from zero, with code amplitudes proportional to the square root of binomial coefficients.  We define the rational number $u = \frac {m}{gn} \ge 1$ to be a scaling factor that determines the length of our quantum code. We call our code with parameters $g,n$ and $u$ a $(g,n,u)$-PI code, or simply, a {\em gnu} code.
The scaling parameter $u$ is related to the energy distribution of a corrupted codeword in a ferromagnetic Heisenberg model \cite{OuF14}; this is beyond the scope of our paper. 

In the logical computation basis $\{ |0_L \> , |1_L\> \}$,
the logical zero and logical one states are
\begin{subequations}
\begin{align}
|0_L\> = \frac{|+_L\>+|-_L\>}{2}
=\sum_{\substack{ \ell {\rm\ even} \\ 0\le \ell \le n }}
\sqrt{ \frac { \bi n \ell} {2^{n-1} } } |D_{g \ell}^{gnu} \>
,\\
|1_L\>=\frac{|+_L\>-|-_L\>}{2}
=\sum_{\substack{ \ell {\rm\ odd} \\ 0\le \ell \le n }}
\sqrt{ \frac { \bi n \ell} {2^{n-1} } } |D_{g \ell}^{gnu} \>
,\notag
\end{align}
\end{subequations}	
and are supported on the Dicke states with excitation numbers $g\ell$ for even $\ell$ and odd $\ell$ respectively. 
Hence gnu codes have their logical states alternately occupy Dicke states of higher excitation number spaced $g$ apart, with maximum occupied excitation number $gn$.

Intuitively, our gnu codes are similar to the harmonic oscillator codes of Gottesman, Kitaev and Preskill \cite{GKP01}. In the limit of infinite $n$, any gnu code is approximately equivalent to an appropriately chosen subspace of a finitely squeezed harmonic oscillator code, because the binomial weightings on the Dicke states on the gnu code approach the Gaussian weightings of the oscillator code. As such, certain limits of our gnu codes may be interpreted as discretized analogues of certain limits of the continuous variable codes of Gottesman, Kitaev and Preskill.

Since we do not expect permutation-invariant code that correct a non-trivial number of errors to be quantum stabilizer codes \cite{gottesman-thesis}, in this paper we introduce techniques from combinatorics and operator theory for analyzing permutation-invariant codes.
Theorem \ref{thm:PI-sparse} and Theorem \ref{thm:AD} quantify the performance of our gnu codes with respect to sparse errors and spontaneous decay errors respectively.
In particular, we prove that (i) if $g=t+1$, $n>3t$ and 
$u \ge 1 +  \frac{t} {gn} $, the gnu code corrects $t$ spontaneous decay errors. An example is the $(2,4,1+\frac{1}{8})$-PI code with logical codewords
\begin{subequations}
\begin{align}
|0_L\> &= \frac{|D^9_0 \> + \sqrt 6 |D^9_4 \>+ |D^9_8 \>  } {\sqrt{8}}, \\
|1_L\> &= \frac{ \sqrt 4 |D^9_2 \>+ \sqrt 4 |D^9_6 \>  } {\sqrt{8}} . \notag
\end{align}
\end{subequations}
We also prove that (ii) if $g=n=2t+1$ and $u \ge1$, gnu codes correct arbitrary $t$ qubit errors.
Our $(3,3,1)$-PI code is precisely Ruskai's 9-qubit PI code that corrects an arbitrary single qubit error \cite{Rus00}, with logical codewords
\begin{subequations}
\begin{align}
|0_L\> &= \frac{|D^9_0 \> + \sqrt 3 |D^9_6 \>  } {\sqrt{4}}, \\
|1_L\> &= \frac{ \sqrt 3 |D^9_3 \>+ |D^9_9 \>  } {\sqrt{4}} . \notag
\end{align}
\end{subequations}
An example of our extension of Ruskai's 9-qubit PI code is a $(5,5,1)$-PI code that corrects arbitrary single and double qubit errors with logical codewords
\begin{subequations}
\begin{align}
|0_L\> &= 
\frac{|D^{25}_0 \> + \sqrt 10 |D^{25}_{10} \>  + \sqrt 5 
|D^{25}_{20} \>  } {\sqrt{16}}, \\
|1_L\> &= 
\frac{ \sqrt 5 |D^{25}_5 \> + \sqrt 10 |D^{25}_{15} \>  + 
|D^{25}_{25} \>  } {\sqrt{16}} . \notag
\end{align}
\end{subequations}
The combinatorial methods of Section \ref{sec:combinatorics} play a crucial role in the proof of both results, and also implicitly explain the origin of the binomial coefficients in the probability amplitudes of our logical codewords.

Our analysis of gnu codes with respect to spontaneous decay errors requires additional tools, and is hence much more involved than our analysis on sparse errors. While the Knill-Laflamme error correction criterion \cite{KnL97} can be used directly for our analysis on sparse errors, it does not apply directly to our analysis on spontaneous decay errors. 
Hence we supply a 
generalization of both the Knill-Laflamme quantum error correction criterion \cite{KnL97} and the approximate quantum error correction criterion of Leung, Nielsen, Chuang and Yamamoto \cite{LNCY97} -- Theorem \ref{thm:pkl}.
Using only the trace, the {\em total deviation}, 
and the smallest eigenvalue of our matrix of code-expectations, 
we quantify the performance of any quantum code with respect to any known noisy process. 
Our work uses purely algebraic means to extend our knowledge of non-stabilizer codes of which many are topological \cite{Kit03}, as opposed to optimization techniques \cite{Kosut08,Fletcher08,Yam09,TKL10} among other approaches \cite{LangShor07,SSSZ09,DGJZ10}. 
We prove Theorem \ref{thm:pkl} by repeatedly applying the Ger\v sgorin circle theorem (see Theorem \ref{thm:GCT}, \cite{Ger31,varga-GCT}). 

This paper has the following structure: In Section \ref{sec:quant-code-perf}, we introduce notations related to our gnu codes, including quantum channels, quantum codes, worst case errors, and $t$-sparse channels. 
In Section \ref{sec:combinatorics}, we introduce our combinatorial lemmas that are crucial in our analysis of gnu codes, the most important of which is Lemma~\ref{lem:cute-binomial-identity}.
In Section \ref{sec:sparse-qecc}, we prove that $(2t+1,2t+1,1)$-PI codes can correct arbitrary $t$ qubit errors (we call these $t$-sparse errors) in Theorem \ref{thm:PI-sparse}.
In Section \ref{sec:prelims}, we review the truncated recovery map of Leung, Nielsen, Chuang and Yamamoto \cite{LNCY97} and basic definitions and results in matrix analysis that are required in this paper.
In Section \ref{sec:deviation-matrices}, we introduce our deviation matrices from which simple upper bounds on the worst case error can be derived (see Theorem \ref{thm:pkl}).
In Section \ref{sec:AD-qecc}, we prove that gnu codes can be used to correct multiple spontaneous decay errors in Theorem \ref{thm:AD}.
Finally in Section \ref{sec:discussions}, we discuss the implications of our findings. The reader that wishes to skip our analysis on spontaneous decay errors may omit reading Sections \ref{sec:prelims}, \ref{sec:deviation-matrices}, and \ref{sec:AD-qecc}.

\section{Quantifying code performance} \label{sec:quant-code-perf}
Here our density matrices are always finite dimensional. 
Let a {\em channel} $\mathcal A$ be a linear map from density matrices to density matrices admitting a (non-unique) Kraus decomposition \cite{nielsen-chuang}
\begin{align}
{\mathcal A}(\rho) = \sum_{\tA \in \fK_{\mathcal A} } \tA \rho \tA^\dagger,
\label{eq:kraus-rep}
\end{align}
where $\sum _{ \tA \in \fK_{\mathcal A} } \tA ^\dagger \tA$ evaluates to the identity operator $\mathbb 1$. 
We call $\fK_{\mathcal A}$, a set of complex matrices, a {\em Kraus set} of ${\mathcal A}$. Elements of a Kraus set are called {\em Kraus operators} or {\em effects} \cite{kraus}, and we call any strict subset of $\fK_{\mathcal A}$ a {\it truncated Kraus set} of ${\mathcal A}$ \cite{OuN13}. Truncating the Kraus set in Eq.~(\ref{eq:kraus-rep}) may cause ${\mathcal A}$ to no longer preserve trace. 
In this case, the operator $\mathbb 1 - \sum _{ \tA \in \fK_{\mathcal A} } \tA ^\dagger \tA$ is positive semidefinite and need not evaluate to zero.
Such truncated maps are also called quantum operations \cite{nielsen-chuang}. 
If the equality
\[
\sum_{\tA \in \Omega } 
w_{\tA,\tF} w_{\tA,\tE}^* 
= \delta_{\tE,\tF}\]
holds for all 
$\tE,\tF \in \Omega$, 
we call
$f(\tA) := \sum_{\tF \in \Omega} w_{\tA, \tF} \tF$
a {\em transformed Kraus operator} 
because for all $\rho$,
\begin{align}
\sum_{\tA \in \Omega} \tA \rho \tA^\dagger 
= \sum_{\tA \in \Omega} f(\tA) \rho f(\tA)^\dagger. \label{eq:transformed-Kops}
\end{align}
Eq.~(\ref{eq:transformed-Kops}) explains the non-uniqueness of the Kraus representation of channels in Eq.~(\ref{eq:kraus-rep}).

Mathematically, a {\em code} is a subspace of a complex Euclidean space where quantum information resides \cite{nielsen-chuang}. 
For a quantum operation $\Phi$, we use the {\em entanglement fidelity} $F(\rho,{\Phi})$ \cite{nielsen-chuang} to quantify the closeness between the density matrices $\rho$ and ${\Phi}(\rho)$. In Schumacher's representation (Eq.~(43) of \cite{Sch96}),
\begin{align}
 F(\rho, {\Phi}) = \sum_{\tB \in \fK_{{\Phi}}} \left|\tr (\tB \rho   ) \right|^2,\notag
\end{align}
where $\tr$ is the matrix trace operator.
If  ${\Phi}=\cR \circ \cA$, where $\cR$ is a recovery channel designed to undo the noisy channel $\cA$, we may interpret 
\begin{align}
E_{\cA,\cC}(\cR) := \max_{\rho \in \fD(\cC)} ( 1-  F(\rho, \cR \circ \cA) )
\label{eq:worst-case-error}
\end{align}
as the corresponding {\em worst case error} of a code $\cC$ after implementing the recovery $\cR$.
Here, $\fD(\cC)$ denotes the set of all density matrices $\rho$ such that 
\[\sum_{|\beta\>\in \cB} \<\beta|\rho| \beta\>=1,\]
where $\cB$ is any orthonormal basis of $\cC$.
For example, the worst case error is always an upper bound on the probability of a logical bit or phase flip. 

Consider the set of {\em Pauli errors} on $m$ qubits, which we denote as 
$\{ \tI, \tX, \tY, \tZ\} ^{\otimes  m}$, where 
$\tI =  \begin{pmatrix}
1 & 0 \\
0 & 1 \\
\end{pmatrix},$ 
$\tX =  \begin{pmatrix}
0 & 1 \\
1 & 0 \\
\end{pmatrix}, $
$\tZ =  \begin{pmatrix}
1 & 0 \\
0 & -1 \\
\end{pmatrix}, $ and 
$\tY =  i \tX \tZ$ are the usual Pauli matrices.
Given a Pauli error $\tP$,  we define its {\em weight} $\wt(\tP)$ to be the number of qubits it acts non-trivially on.
We say a linear combination of Pauli errors is $t$-{\em sparse} if each of the constituent Pauli operators with a non-zero coefficient has a weight no greater than $t$.
We say a quantum channel is $t$-sparse if each of its Kraus operator is also $t$-sparse.
We prove in Theorem \ref{thm:PI-sparse} that given any $t$-sparse channel $\mathcal A$ that acts on a single qubit encoded in our $(2t+1,2t+1,u)$-PI code $\mathcal C$ for all feasible scaling factors $u \ge 1$, there exists a recovery channel $\mathcal R$ for which the worst case error is exactly zero, that is
\begin{align}
  E_{\cA, \cC}(\cR) = 0 \label{eq:sparse-perfect-error-correction}.
\end{align}
In general, the problem of error correction is also equivalent to the `min-max' problem
\begin{align}
\inf_\cR  E_{\cA, \cC}(\cR)  = \inf_\cR \max_{\rho \in \mathfrak D(\cC)} (1 - F(\rho, \cR \circ \cA)), \notag
\end{align}
where we choose the best recovery channel $\cR$ for the worst case density matrix $\rho$ in our codespace.

A phenomenological model for the spontaneous decay of `probability' $\gamma$ on a two-level system is the {\em amplitude damping} (AD) channel $\cA_\gamma$, with Kraus operators 
\begin{align}
\tA_0 = |0\>\<0| +\sqrt{1-\gamma} |1\>\<1|, \quad \tA_1 = \sqrt{\gamma} |0\>\<1|. \label{eq:AD-definition}
\end{align}
If the channel $\cA_\gamma$ accurately describes experimentally observed decoherence, the equation $1-\gamma = e^{-\tau/T_1}$ implicitly quantifies $\gamma$ in terms of an experimentally observed spin-lattice relaxation time ($T_1$) after a time elapse of $\tau$. A Taylor approximation for small $\frac{\tau}{T_1}$ yields
$\gamma \approx	\frac{\tau}{T_1}$.

When there is no encoding of the qubit, the code is just $\mathbb C_2$ and hence the error which corresponds to no recovery is $E_{\cA_\gamma, \bbC^2}(\cI) = \gamma$, where $\cI$ is the identity channel on a qubit.
If one encodes a qubit into a quantum code over multiple qubits, it might be possible to mitigate the effects of amplitude damping erros, in the sense that for some positive integer $t$, our min-max problem for error correction has for all $ \gamma \in [0, \gamma_0]$ the upper bound
\begin{align}
\inf_\cR E_{\cA_{\gamma}^{\otimes m}, \cC}(\cR) \le C\gamma^{t+1}, \label{eq:tAD-def}
\end{align}
for some positive constants $C$ and $\gamma_0$.
Codes for which Eq.~(\ref{eq:tAD-def}) hold are called $t$-amplitude damping codes, or $t$-AD codes for short.
By definition, a $t$-AD code suppresses the error probability $\gamma$ by $t$-folds in the exponent. 
We prove in Theorem \ref{thm:AD} that a gnu code with 
$g = t+1$, $n>3t$, and $u \ge 1 + \frac{t}{gn}$
is a $t$-AD code for all positive integers $t$~(see Fig.~\ref{fig:1}).


\section{Combinatorics} \label{sec:combinatorics}
In this section we introduce the key combinatorial results that we use to prove our gnu codes' utility in combating both amplitude damping and sparse errors. 

Denote the binomial coefficient as $\bi{n}{\ell} := \frac{n_{(\ell)}}{\ell!}$ where
the falling factorial is 
\[{n_{(\ell)} :=\prod_{k=0}^\ell (n-k)}.\]
Our main combinatorial tool is the following lemma.
\begin{lemma}
\label{lem:cute-binomial-identity}
Let $n$ be a positive integer. Then for all integers $x$ such that 
$0\le x \le n-1$, 
\begin{align}
\sum_{\ell=0}^{n} \bi{n}{\ell} \ell^x (-1)^\ell = 0.
\label{lem:pkl-combinatorial}
\end{align}
\end{lemma}
The above lemma can be proved trivially using linear combinations of the binomial identity (Eq. (11) in Page 609 of \cite{PBM86})
\begin{align}
\sum_{\ell=0}^{n} \bi{n}{\ell} \ell_{(x)} (-1)^\ell = 0, \notag
\end{align}
which holds for all $0 \le x < n$, and can be proved by considering the derivatives of the binomial generating function and using induction.

The lemma below represents a fraction of binomial coefficients $\frac{  \bi { m - (a + c) }{ w - a }  }
	 {  \bi { m}  { w } } 
$ 
as a polynomial in $w$.
\begin{lemma}
\label{lem:binom--frac-to-polynomial}
Let $a, c, m $ and $w$ be non-negative integers such that 
$a \le w \le m-c$. Then 
\[ 
\frac{  \bi { m - (a + c) }{ w - a }  }
	 {  \bi { m}  { w } } 
= \bi{w}{a} \frac{a!(m-w)_{(c)}}{m_{(a+c)}}.
\]
\end{lemma}
\begin{proof}
Let $s = m - (a+c)$.
Then,
\begin{align}
\frac{\bi{s}{w-a}}{\bi{m}{w} }
&= \frac{s_{(w-a)} w! a!}{m_{(w)}(w-a)!a!} 
= \bi{w}{a} \frac{a! s_{(w-a)}}{m_{(w)}}. \notag
\end{align}
We can rewrite the above equation using the identity
\[
\frac{s_{(w-a)}}{m_{(w)}} 
= \frac{(m-a-c)_{(w-a)}}{m_{(w)}}
= \frac{(m-w)_{(c)}}{m_{(a+c)}}
\]
to get 
\[
\frac{\bi{s}{w-a}}{\bi{m}{w} } 
= \bi{w}{a} \frac{a!(m-w)_{(c)}}{m_{(a+c)}}.
\]
\end{proof}
To study the correction of spontaneous decay errors, we consider the following.
\begin{lemma}
\label{lem:inner-product-series-representations-taylor}

Let $a, c, m $ and $w$ be non-negative integers as given by Lemma \ref{lem:binom--frac-to-polynomial}. Then 
${\gamma^{a} 	(1-\gamma)^{w-a}
    \frac{\bi{  m - (a+c)}{w-a}}{\bi{m}{w}}} $ 
has the Taylor series
\begin{align}
\sum_{k=a}^{m}
	 \frac{(-1)^{y}   k_{(a)} }
	 {m_{(a+c)}}
		  (m-w)_{(c)}
         \bi{w}{k} \gamma^{k}. 
\label{eq:matrix-elements-decompositions-h}
\end{align}
\end{lemma}
\begin{proof}
Using Lemma \ref{lem:binom--frac-to-polynomial}, the binomial expansion 
\[ (1-\gamma)^{w-a} 
= \sum_{y=0}^{w-a}
	 \bi{w-a}{y}(-1)^y \gamma ^{y}, \]
and the identity
\[\bi{w-a}{y}\bi{w}{a} 
= \frac{w_{(a)} (w-a)_{(y)}   }{a! y!}
=\bi{w}{a+y} \bi{a+y}{a},
\]
and setting $k=a+y$ yields the result.
\end{proof}

\section{Correcting sparse errors}
\label{sec:sparse-qecc}
In this section, we consider gnu codes with gap $g=2t+1$, occupation number $n=2t+1$, and scaling factor $u \ge 1 $, for all positive integers $t$.
We investigate the utility of our PI codes in protecting an encoded qubit from $t$-sparse errors.

Let ${\mathcal A}$ be a  $t$-sparse channel with Kraus set $\Omega$. 
Note that for all $\tA, \tB \in \Omega$, the matrix $\tA ^\dagger \tB$ has a maximum weight of $2t$.
Hence for our code analysis, it suffices to evaluate inner products of the form
\begin{align}
\< 0_L | \tA ^\dagger \tB  | 0_L \>, \quad
\< 1_L | \tA ^\dagger \tB  | 1_L \>, \mbox{ and }
\< 0_L | \tA ^\dagger \tB  | 1_L \>. \notag
\end{align}
Clearly the cross-term $\< 0_L | \tA ^\dagger \tB  | 1_L \>$ is zero for all $\tA, \tB \in \Omega$, because our gap $g=2t+1$ is strictly greater than the maximum weight of $\tA ^\dagger \tB$. 
Indeed, if 
\begin{align}
\< 0_L | \tA ^\dagger \tB  | 0_L \> - \< 1_L | \tA ^\dagger \tB  | 1_L \> 
\label{eq:sameness}
\end{align}
equals to zero for all $\tA, \tB \in \Omega$, 
the Knill and Laflamme quantum error correction conditions \cite{KnL97} will hold, and perfect correctibility will follow. In this section we prove that our $(g,n,u)$-PI code corrects perfectly with respect to all $t$-sparse noisy channels.
\begin{theorem}
\label{thm:PI-sparse}
Let $t$ be a positive integer and $\Omega$ be the Kraus set of any $t$-sparse channel.
Then the worst case error is exactly zero with respect to a gnu code where $g=n=2t+1$, and $u \ge 1 $.
\end{theorem}
As a first step to prove Theorem \ref{thm:PI-sparse}, observe that (\ref{eq:sameness}) simplifies to 
\begin{align}
2^{n-1} \sum_{0 \le \ell \le n} (-1)^\ell   \bi{n}{\ell } 
\< D^m_{g \ell} | \tA ^\dagger \tB | D^m_{ g \ell } \>. \notag
\end{align}

\begin{lemma} 
\label{lem:PI-sameness}
Let $g,n$ and $u$ be as defined in Theorem \ref{thm:PI-sparse}, and let $m = gnu. $
For any Pauli error $\tP$ in $\{ \tI, \tX, \tY, \tZ \} ^{\otimes m}$ of weight strictly less than $n$, we have
\begin{align}
&\<0_L| \tP| 0_L\> - 
\<1_L| \tP| 1_L\> \notag\\
=&
\sum_{0 \le \ell \le n} (-1)^\ell   \bi{n}{\ell } 
\< D^m_{g \ell } | \tP | D^m_{ g \ell } \>= 0. \notag
\end{align}
\end{lemma}
\begin{proof}
The first equality in the lemma is obvious by definition.
The permutation invariance of the Dicke states allows us to assume without any loss of generality that 
\[ 
	\tP = \tE \otimes \tI ^{\otimes m- (x+y+z)}
\]
where 
\[\tE = \tX^{\otimes x} \otimes \tY ^{\otimes y} \otimes \tZ ^{\otimes z},\]
and $x,y$ and $z$ are non-negative integers such that 
\[x+y+z = \wt(\tP).\]
For $w = g \ell$, let $B^m_w$ be the set of all binary vectors of length $m$ and weight $w$.
Consider the set analogue of the Vandermonde identity
\[\bi{m}{w} = \sum_{a=0}^{\wt(\tP)} \bi{\wt(\tP)}{a} \bi{m-\wt(\tP)}{w-a},\] which 
decomposes a set of constant weight vectors into the following union of Cartesian products of sets:
\begin{align}
B^m_w = \bigcup _{a = 0}^{\wt(\tP)} B^{\wt(\tP)}_a \times B^{m-\wt(\tP)}_{w-a}. \notag
\end{align}
Then it follows that 
\begin{align}
&\bi{m}{w } \<D^m_w | \tP | D^m_w \> =
\sum_{{\bf x}, {\bf y} \in B^m_w } \< {\bf x}| \tP  |{\bf y}\> \notag\\
=&
\sum_{a=0}^{\wt(\tP)} 	\sum_{{\bf a}, {\bf b} \in B^{\wt(\tP)}_a } 
\< {\bf a}| \tE  |{\bf b}\> \bi{m-\wt(\tP)}{w-a }. \notag
\end{align}
To simplify the expression on the right hand side of the above equation,
define the function
\[
	f(\tE,a,\wt(\tP)) :=\sum_{{\bf a}, {\bf b} \in B^{\wt(\tP)}_a } 
\< {\bf a}| \tE  |{\bf b}\>.
\]
Hence
\begin{align}
&\sum_{0 \le \ell \le n} (-1)^\ell   \bi{n}{\ell } 
\< D^m_{g \ell} | \tP | D^m_{g \ell } \> \notag\\
=&
\sum_{0 \le \ell \le n} (-1)^\ell   \bi{n}{\ell } 
\sum_{a=0}^{\wt(\tP)} 	f(\tE,a,\wt(\tP)) 
\frac{
	\bi{m-\wt(\tP)}{g \ell-a }
}{ \bi{m} {g \ell }}. \label{eq:area51}
\end{align}
Exchanging the order of summation in (\ref{eq:area51}) and noting that $f(\tE,a,\wt(\tP))$ does not depend on $\ell$, we get
\begin{align}
\sum_{a=0}^{\wt(\tP)} f(\tE,a,\wt(\tP)) 
\left(	 \sum_{0 \le \ell \le n} (-1)^\ell   \bi{n}{\ell } 
\frac {  \bi{m-\wt(\tP)}{g \ell-a } }  { \bi{m}  {g \ell } } \right).
\label{eq:pauli-sparse}
\end{align}
Using Lemma \ref{lem:binom--frac-to-polynomial}, the ratio of binomial coefficients in 
(\ref{eq:pauli-sparse}) is 
a polynomial in $\ell$ of order $\wt(\tP)$ given by
\[
\frac{
	\bi{m-\wt(\tP)}{g \ell-a }  
}{ \bi{m} {g \ell }}
 = 
\bi{g \ell }{a} \frac{a! (m-\wt(\tP) )_{(\wt(\tP) - a)} } { m_{(\wt(\tP) )}}.
\]
Hence Lemma \ref{lem:cute-binomial-identity} and the inequality $n > \wt(\tP)$ imply that the bracketed term in (\ref{eq:pauli-sparse}) is zero, which proves the result.
\end{proof}
The above lemma implies that the expression in (\ref{eq:sameness}) is always zero for all Pauli errors $\tA$ and $\tB$ of both of weight no more than $t$, and hence Theorem \ref{thm:PI-sparse} follows from the Knill-Laflamme error correction criterion. 
For completeness, we prove Theorem \ref{thm:PI-sparse} formally in the last part of Section \ref{sec:deviation-matrices}.

We remark that our $(3,3,1)$-PI code on 9 qubits is precisely Ruskai's {9-qubit} PI code \cite{Rus00} that corrects an arbitrary single qubit error. This $(3,3,1)$-PI code is in fact, a completely symmetrized version of the Shor code \cite{ShorCode}, as noted by Ruskai \cite{Rus00}. Similarly our 
{$(2t+1,2t+1,1)$-PI} codes are just completely symmetrized versions of the Bacon-Shor codes \cite{Bac06,A-subsys} that are capable of correcting arbitrary $t$-sparse errors.

\section{A Review of Truncated recovery maps and matrix analysis}
\label{sec:prelims}
\subsection{Truncated recovery maps}
We now review the recovery map of Leung, Nielsen, Chuang and Yamamoto \cite{LNCY97}. Given a basis $\mathcal B$ of a code $\mathcal C$, let $\Pi = \sum_{|\beta\>\in\cB} |\beta\>\<\beta|$ be a projector of states into the codespace $\cC$.
For all $\tA$ in the truncated Kraus set $\Omega$,
define \[\Pi_{\tA} := \tU_\tA \Pi {\tU_\tA}^\dagger,\]
where $\tU_\tA$ is the unitary
in the polar decomposition $\tA \Pi = \tU_\tA \sqrt{\Pi \tA^\dagger \tA \Pi}$. 
Also define the recovery operator 
\[\tR_\tA := \tU_\tA^\dagger \Pi_{\tA}.\]
One might hope to use the {\it truncated recovery map} of Leung {\em et al.}
\begin{align}
\mathcal R_{\Omega, \cC}(\mu) := \sum_{\tA \in \Omega} \tR_\tA \mu \tR_\tA^\dagger, \notag
\end{align}
for code recovery,
because it can be implemented by performing a projective measurement followed by applying a unitary which depends on the previous measurement outcome.
Non-orthogonal projectors $\Pi_\tA$ however cause 
$\mathcal R_{\Omega, \cC}$ to increase trace and not be a quantum operation.

Regardless of whether $\mathcal R_{\Omega,\mathcal C}$ is a quantum operation, using Schumacher's formula for the entanglement fidelity (equation (43) of \cite{Sch96}) and omitting terms of the form $|\tr (\tR_\tA \tB \rho)|^2$ for distinct $\tA, \tB \in \Omega$ formally gives
\begin{align}
\mathcal F_e(\rho, \mathcal R_{\Omega, \cC} \circ {\mathcal A})
\ge&
\sum_{\tA \in \Omega}
| \tr ( \tR_{\tA} \tA \rho )|^2.  \notag
\end{align}
The expression on the right hand side of the above inequality admits the following lower bound:
\begin{lemma}[Leung, Chuang, Nielsen, Yamamoto \cite{LNCY97}]
\label{lem:leung1}
	\begin{align}
	&\sum_{\tA \in \Omega} \left| \tr (\tR_{\tA}  \tA \rho)    \right|^2 \ge
	\sum_{\tA \in \Omega} \lambda_{\min}(\tA ^\dagger \tA) \notag
	\end{align}
\end{lemma}
\subsection{Matrix Analysis}
For a $d$-dimensional complex vector ${\bf v} = \{v_1, \dots, v_d\}$ and real number $p$ such that $p\ge 1$, define the vector $p$-norm of ${\bf v}$ as 
\begin{align}
\| {\bf v} \|_p  := \left(
\sum_{j=1 }^d v_j^p 
\right)^{1/p}.\notag
\end{align}
Let $\tG : \bbC^d \to \bbC^d$ be a finite dimensional linear map.
The linear map $\tG$ can be represented as a finite dimensional square matrix, and we define its operator $p$-norm as
\begin{align}
 \| \tG  \|_p : = \sup \{ \| \tG {\bf v}  \|_p : 
	{\bf v}  \in \bbC^d , \| { \bf v}  \|_p = 1  \} .\notag
\end{align}
If $\tG$ is a positive semidefinite matrix, its operator 2-norm is just its spectral radius (and maximum eigenvalue), and 
\begin{align}
	\| \tG ^n \|_2 
= 	(\| \tG \|_2)^n \label{eq:2norm-spectral}
\end{align}
for all positive integers $n$.

When the off-diagonal elements of a square complex matrix $\tG$ are vanishingly small, Ger\v sgorin's classic result \cite{Ger31,varga-GCT}
approximates the eigenvalues of $\tG$ using its diagonal entries.
\begin{theorem}[Ger\v sgorin circle theorem \cite{Ger31,varga-GCT}]
\label{thm:GCT}
Let $g_{i,j}$ be the matrix elements of a $d \times d$ matrix $\tG$,
with row index $i$ and column index $j$ chosen from $[d]$. 
Then every eigenvalue of $\tG$ lies within the union of the the Ger\v sgorin discs $D_i$, where
\[
D_i := \{ x \in \bbC :  |x - g_{i,i}| \le \sum_{j\neq i } | g_{i,j}|   \} .
 \]
\end{theorem}

\section{Deviation matrices and quantum error correction}
\label{sec:deviation-matrices} 
Given a channel with a truncated Kraus set $\Omega$, and a code with orthonormal basis $\cB$, for each $\tA$ and $\tB$ in $\Omega$ we may evaluate the code-averaged expectations
\begin{align}
g_{\tA, \tB}  := \frac{1}{ |\cB| } \sum_{|\b\> \in \cB } \< \b |\tA^\dagger \tB | \b \>.\notag
\end{align}
We rearrange the code-averaged expectations $g_{\tA,\tB}$ into a matrix $\tG$ and its corresponding ($\tA,\tB$)-deviation matrices defined respectively by
\begin{subequations}
\begin{align}
\tG &:= \sum_{\tA,\tB \in \Omega} g_{\tA,\tB} |\tA\>\<\tB|,
\label{eq:Gmatrix}\\
\tG_{\tA, \tB}  &:= \!\!\!
	\sum_{|\alpha\>, |\beta\> \in \cB} \!\!\!
		(\<\alpha| \tA ^\dagger \tB |\beta\> - 
		 g_{\tA,\tB} \d_{|\alpha\>,|\beta\>}) |\alpha\>\<\beta| \label{eq:perturbation},
\end{align} 
\end{subequations}
where the orthonormal basis $ \{ |\tE\> : \tE \in \Omega \}$ labels the Kraus operators in $\Omega$. 
Each deviation matrix $\tG_{\tA, \tB}$ has a diagonal and an off-diagonal matrix element of maximal magnitude, 
which we denote as $\theta_{\tA,\tB}$ and $\sigma_{\tA,\tB}$ respectively. 
Define the {\em total deviation}
\begin{align}
\epsilon := \max_{\tA,\tB} \theta_{\tA,\tB} +
(|\cB|-1) \max_{\tA,\tB} \sigma_{\tA,\tB}.\label{eq:deviations}
\end{align}
The total deviation $\epsilon$, $\tr\tG$,
and the minimum eigenvalue of $\tG$ are the only ingredients of Theorem \ref{thm:pkl}. 

We give a lower bound on the magnitude of the rescaling factor $\eta$ for which our truncated recovery map $\cR_{\Omega, \cC, \eta} := \frac  1 {1+\eta} \cR_{\Omega,\cC}$ is a valid quantum operation:
\begin{lemma}
\label{lem:rescaled-map}
Let $\eta$ be a non-negative real number such that
$
\eta \ge \sum_{\tA \neq \tB \in \Omega} \left\| \Pi \tU_\tA^\dagger \tU_{\tB} \Pi \right\|_2. 
$
Then the map $\cR_{\Omega, \cC, \eta}$ is a quantum operation.
\end{lemma}
\begin{proof}
It suffices to show that 
\[ \frac{1}{1+\eta}
 \left \|  \sum_{\tA \in \Omega } \tR_\tA^\dagger \tR_\tA  
 \right \|_2 \le 1 . \]
First note that $\tR_\tA^\dagger \tR_\tA = 
(\Pi_\tA ^\dagger \tU_\tA)( \tU_\tA ^\dagger \Pi_\tA ) = \Pi_\tA$. 
Since the projectors $\Pi_{\tA}$ may not be orthogonal, 
\begin{align}
\left( \sum_{\tA \in \Omega} \tR_\tA^\dagger \tR_\tA \right)^2 
&= 
\Bigl( \sum _{\tA \in \Omega} \Pi_{\tA} \Bigr) \Bigl( \sum _{\tB \in \Omega} \Pi_{\tB} \Bigr) \notag\\
&=
   \sum_{\tA \in \Omega } \Pi_{\tA}  
     +   \sum_{\tA \neq \tB \in \Omega } \tU_\tA \Pi \tU_\tA^\dagger \tU_{\tB} \Pi \tU_\tB^\dagger. \label{eq:expand-projectors}
\end{align}
Since the left hand side of the above equation is a positive semidefinite matrix, we use (\ref{eq:2norm-spectral}) to get
\begin{align}
  \Bigl \| \Bigl( \sum_{\tA \in \Omega} \tR_\tA^\dagger \tR_\tA \Bigr)^2 \Bigr \|_2 &=
  \Bigl \| \sum_{\tA \in \Omega} \tR_\tA^\dagger \tR_\tA  \Bigr \|_2^2.\notag
\end{align}
Applying the operator 2-norm on both sides of (\ref{eq:expand-projectors}), using the triangle inequality for operator norms with the above inequality, and applying the unitary invariance of the operator 2-norm then gives
\begin{align}
\left \| \sum_{\tA \in \Omega} \tR_\tA^\dagger \tR_\tA \right \|_2^2 
&\le  
\left\|   \sum_{\tA \in \Omega } \Pi_{\tA}  \right\|_2
     +   \sum_{\tA \neq \tB \in \Omega } \| \Pi \tU_\tA^\dagger \tU_{\tB} \Pi \|_2 \notag\\
&\le  
\left\|   \sum_{\tA \in \Omega } \Pi_{\tA}  \right\|_2
     +   \eta. \label{eq:almost-there-inequality}
\end{align}
Define $\theta$ to be a real number such that 
\[
\left\|   \sum_{\tA \in \Omega } \Pi_{\tA}  \right\|_2 = 1+ \theta.
\]
Since the operator 2-norm of each of the projectors $\Pi_\tA$ is at least one, the real number $\theta$ has to be non-negative.
Now the inequality (\ref{eq:almost-there-inequality}) is equivalent to 
\[
(1+\theta)^2  \le (1+ \theta) +  \eta. 
\]
The above inequality is equivalent to 
\[
1+\theta + \theta^2  \le 1+ \eta. 
\]
Hence it follows that $(1+\theta)  \le (1+ \eta). $ Applying the definition of $\theta$ then gives 
\[
\left \|  
	\sum_{\tA \in \Omega} \tR_\tA ^\dagger \tR_\tA
\right\|_2 \le 1+\eta,
\]
and the result follows.
\end{proof}

The code's error using the rescaled recovery $\cR_{\Omega, \cC, \eta}$ is
\begin{align}
E_{\cA,\cC}(\cR_{\Omega,\cC,\eta}) \le   1- \sum_{\tA \in \Omega} \frac{\lambda_{\min,\cC} ( \tA^\dagger \tA   )}{1+\eta}, 
\label{eq:trun-rec-err-bound}
\end{align}
where $\lambda_{\min,\cC}(\cdot)$ denotes the minimum eigenvalue of a matrix restricted to a subspace $\cC$.
The bound in (\ref{eq:trun-rec-err-bound}) follows from trivial application of Lemma \ref{lem:leung1} and Lemma \ref{lem:rescaled-map}, and with the definition of the worst case error given by (\ref{eq:worst-case-error}).

When the projectors $\Pi_\tA$ are orthogonal, we may set $\eta=0$ in the bound (\ref{eq:trun-rec-err-bound}) to recover the result of Leung, Nielsen, Chuang, and Yamamoto regarding an approximate quantum error correction criterion \cite{LNCY97}.

Motivated by Knill and Laflamme's methodology \cite{KnL97}, 
we consider the spectral decomposition of the Hermitian matrix $\tG$;
there exists a unitary matrix $\tV$ such that
$\tD 	:= \tV \tG \tV^\dagger$ is diagonal. 
We use the decompositions
\begin{align}
		\tV 		= \sum_{\tE, \tF \in \Omega} v_{\tE, \tF} |\tE\>\<\tF|, 
		\quad \tD = \sum_{\tE \in \Omega} d_\tE|\tE\>\<\tE|.
\label{eq:diagonal-D}
\end{align}
For all Kraus operators $\tA \in \Omega$, let 
$\widetilde \tA := \sum_{\tF \in \Omega} v_{\tA, \tF} \tF$
denote transformed Kraus operators.
Repeatedly using the Ger\v sgorin circle theorem 
yields the deviation bounds:
\begin{lemma}
\label{lem:pkl-error-bound}
For all distinct $\tA, \tB \in \Omega$, 
\begin{subequations}
\begin{align}
\| \Pi \widetilde \tA ^\dagger \widetilde \tA \Pi	\|_2
\ge d_\tA, \quad 
\| \Pi \widetilde \tA ^\dagger \widetilde \tB \Pi	\|_2
\le |\Omega| \epsilon
\label{ineq:AB_upperbound} \\
\sum_{\tA \in \Omega} \lambda_{\min, \cC}(\widetilde \tA ^\dagger \widetilde \tA ) \ge \tr \tG -  |\Omega| ^2  \epsilon
.\label{ineq:LAA_lowerbound} 
\end{align}
\end{subequations}
\end{lemma}
\begin{proof}
The decompositions in (\ref{eq:diagonal-D}) imply that for all 
$\tA , \tB \in \Omega$, 
\begin{align}
\sum_{\tF, \tF' \in \Omega} 
	v_{\tB,\tF} g_{\tF,\tF'}  v_{\tA,\tF'}^* 
= d_{\tA}\delta_{\tA,\tB}. 
	\label{Seq:unitary-identity}
\end{align}
Substituting (\ref{eq:perturbation}) and (\ref{Seq:unitary-identity}) into 
\begin{align}
\<\alpha | \widetilde \tA^\dagger \widetilde \tB |\beta \>  = \sum_{\tF, \tF' \in \Omega}\<\alpha|\tF ^\dagger \tF' |\beta\>v^*_{\tA, \tF} v_{\tB, \tF'} \notag
\end{align}
gives our version of the `diagonalized and perturbed' Knill-Laflamme conditions \cite{KnL97}
\begin{align}
\<\alpha | \widetilde \tA^\dagger \widetilde \tB |\beta \>  = &d_{\tA} \delta_{\tA, \tB} \delta_{|\alpha\>, |\beta\>} \notag\\ 
& \quad +  \sum_{\tF, \tF' \in \Omega} (v_{\tA, \tF'}^* v_{\tB , \tF} ) 
\< \alpha | \tG_{\tF',\tF} | \beta\>.
\label{Seq:perturbed}
\end{align}
To obtain the first inequality in (\ref{ineq:AB_upperbound}),
observe that 
\[
	\| \Pi \widetilde \tA^\dagger \widetilde \tA \Pi \|_2 
	\ge  \frac{1}{|\mathcal B |} 
	\sum_{|\alpha\> \in \cB} \<\alpha| \widetilde \tA ^\dagger \widetilde \tA | 	\alpha\>. 
\]
Clearly $\displaystyle \sum_{|\alpha\> \in \cB } \<\alpha| \tG_{\tA, \tB}|\alpha\> = 0$, and its substitution into (\ref{Seq:perturbed}) summed over
$|\alpha\>$ with $|\beta\> = |\alpha\>$ gives
\[ \frac{1}{| \cB| }
 \sum_{|\alpha\> \in \cB} 
\<\alpha| \widetilde \tA ^\dagger \widetilde \tA | \alpha\> = d_\tA 
 \ge \lambda_{\min} (\tG). \]
To prove the second inequality in (\ref{ineq:AB_upperbound}),
define the vector 
$\bf v_\tA := (v_{\tA, \tF})_{\tF \in \Omega}$.
Note that 
\[ \sum_{\substack { \tF, \tF' \in \Omega }  }
|v_{\tA, \tF'}^* v_{\tB , \tF} |
= \|   {\bf v}_\tA\|_1 \|  {\bf v}_\tB \|_1.\]
Now the Cauchy-Schwarz inequality implies that 
\[\|   {\bf v}_\tA\|_1 \le \sqrt{|\Omega| \<{\bf v}_\tA , {\bf v}_\tA \>}.\]
Moreover $\bf v_\tA$ is a column vector in the unitary matrix $\tV$ with $\<{\bf v}_\tA, {\bf v}_\tA\> = 1$. Hence applying H\"older's inequality gives
\begin{align}
& \left| \<\alpha | \widetilde \tA^\dagger \widetilde \tB |\beta \> \right| =
\left|\sum
_{\substack { \tF, \tF' \in \Omega  }  } 
v_{\tA, \tF'}^* v_{\tB , \tF}  
\<\alpha| \tG_{\tF',\tF}|\beta \>  \right| \notag\\
&\le 
|\Omega| \max_{\substack { \tF, \tF' \in \Omega  }  }
\left| \<\alpha| \tG_{\tF',\tF}|\beta \>  \right| 
\le \left\{
	\begin{array}{ll} \displaystyle
		|\Omega ||  \sigma_{\tA,\tB} |	&
, |\alpha\> \neq	|\beta\> \\
\displaystyle
		|\Omega ||  \theta_{\tA,\tB}|	&
, |\alpha\> = 		|\beta\> \\
	\end{array}
\right.. \label{Seq:tilde-epsAB-upper-bound}
\end{align}
Applying the Ger\v sgorin circle theorem 
on 
\begin{align}
\sum_{|\alpha\>, |\beta\> \in \cB}|\alpha\>\<\beta| \<\alpha| \widetilde \tA ^\dagger \widetilde \tB | \beta\>
\label{eq:final-perturbation-matrix}
\end{align}
using (\ref{Seq:perturbed}) and (\ref{Seq:tilde-epsAB-upper-bound}) yields the bound.

To prove (\ref{ineq:LAA_lowerbound}), we can 
similarly apply the Ger\v sgorin circle theorem 
on (\ref{eq:final-perturbation-matrix}) with 
$\widetilde \tB = \widetilde \tA$
to get a lower bound on
$\lambda_{\min, \cC}(\widetilde \tA ^\dagger \widetilde \tA)$ which when summed over $\tA \in \Omega$ yields the required lower bound.
\end{proof}

\begin{theorem}\label{thm:pkl}
Let $\Omega$ be a truncated Kraus set, $\cC$ be a code with orthonormal basis $\cB$, and let $\tG$ and $\epsilon$ be as given by Eqs.~(\ref{eq:Gmatrix}) and (\ref{eq:deviations}). If
$\eta = \frac{(|\Omega|-1) |\Omega|^2 \epsilon}{ \lambda_{\min}(\tG)}$,
then 
\begin{align}
\inf_\cR E_{\cA,\cC}( \cR) \le 1- \frac{\tr \tG - |\Omega|^2 \epsilon}{1+ \eta }.\label{eq:thmpkl}
\end{align}
\end{theorem}
To prove Theorem \ref{thm:pkl}, 
we could use recovery maps guaranteed to be nearly optimal \cite{BaK02,Tys10,BeO10,BeO11}, but 
we set $\cR=\cR_{\Omega,\cC,\eta}$ because $\cR_{\Omega,\cC,\eta}$ possibly quantifies the performance of a recovery implemented by Leung~{\em et al.}'s recovery circuit \cite{LNCY97}.
\begin{proof}
The polar decompositions
$\widetilde \tA  \Pi = \tU_{\widetilde \tA} \sqrt{ \Pi \widetilde \tA^\dagger \widetilde \tA\Pi}$ and
$\widetilde \tB  \Pi = \tU_{\widetilde \tB} \sqrt{ \Pi \widetilde \tB^\dagger \widetilde \tB\Pi}$ for distinct $\tA, \tB \in \Omega$ imply that
\begin{align*}
\Pi \widetilde \tA^\dagger \widetilde \tB \Pi
= \sqrt{\Pi \widetilde \tA^\dagger \widetilde \tA \Pi} ( \Pi \tU_{\widetilde \tA}^\dagger \tU_{\widetilde \tB} \Pi )  \sqrt{\Pi \widetilde \tB^\dagger \widetilde  \tB \Pi}. \notag
\end{align*}
Sub-multiplicativity of norms implies that 
\begin{align}
\|\Pi  \tU_{\widetilde \tA}^\dagger \tU_{\widetilde \tB}  \Pi  \|_2 
\le 
\frac{ \| \Pi \widetilde \tA^\dagger \widetilde \tB \Pi \|_2}
{
 	\|   \sqrt{\Pi \widetilde \tA^\dagger \widetilde \tA \Pi} \|_2    
      \| \sqrt{\Pi \widetilde \tB^\dagger \widetilde \tB \Pi}  \|_2 
}
. \label{ineq:submult}
\end{align}  
Using (\ref{ineq:AB_upperbound}), the upper bound in (\ref{ineq:submult}) is at most
\begin{align}
       \frac{\| \Pi \widetilde \tA^\dagger \widetilde \tB \Pi \|_2 }{ \min_{\tF \in \{ \tA, \tB\} }\lambda_{\max} (\Pi \widetilde \tF^\dagger \widetilde \tF \Pi)} 
       \le
       \frac{|\Omega|  \epsilon }{ \lambda_{\min}(\tG) }. \notag
\end{align}
Hence if $\eta \ge 
\frac{|\Omega|^2(|\Omega|-1)  \epsilon }{ \lambda_{\min}(\tG) }$,
Lemma \ref{lem:rescaled-map} holds, implying that $\cR_{\Omega,\cC, \eta}$ is a quantum operation. 
The upper bound (\ref{eq:thmpkl}) comes by substituting $\tA$ with $\widetilde \tA$ and applying (\ref{ineq:LAA_lowerbound}) in (\ref{eq:trun-rec-err-bound}). 
\end{proof}
If the total deviation $\eps$ is zero, Theorem \ref{thm:pkl} is equivalent to Knill and Laflamme's result on perfect quantum error correction \cite{KnL97}.

We now prove Theorem \ref{thm:PI-sparse} by invoking a special case of our main technical result Theorem \ref{thm:pkl}. This illustrates concretely the reduction of Theorem \ref{thm:pkl} to the Knill-Laflamme quantum error correction conditons \cite{KnL97} when all the deviation matrices $\tG_{\tA,\tB}$ are exactly zero.
\begin{proof}[Proof of Theorem \ref{thm:PI-sparse}]
Using Lemma \ref{lem:PI-sameness} with the expression in (\ref{eq:sameness}), it follows that for all $\tA,\tB \in \Omega$, 	the deviation matrices $\tG_{\tA,\tB}$ are identically zero.  Hence it follows from Theorem \ref{thm:pkl} that the 
worst case error is at most
\begin{align}
&1 - \tr (\tG)  \notag \\
=& 1 
-\frac{  \<0_L| \sum_{\tA \in \Omega} \tA ^\dagger \tA |0_L\> 
+ \<1_L| \sum_{\tA \in \Omega} \tA  ^\dagger \tA |1_L\> }{2}.
\label{eq:expr-1-G}
\end{align}
Using the completeness relation $\sum_{\tA \in \Omega} \tA  ^\dagger \tA = \mathbb 1$, the expression (\ref{eq:expr-1-G}) simplifies to yield 
\begin{align}
1  -\frac{  \<0_L| 0_L\> +  \<1_L |1_L\> }{2} = 0. \notag
\end{align}
This implies that noise induced by the channel ${\mathcal A}$ can be perfectly reversed using our gnu code. 
\end{proof}

\section{Correcting spontaneous decay errors}
\label{sec:AD-qecc}

In this section, we consider gnu codes (\ref{eq:PI-code}) 
of length $m=gnu$ with a gap $g = t+1$, occupancy number $n > 3t$, and scaling factor $u \ge 1 + \frac{t} {gn } $ for positive integers $t$.
We consider the amplitude damping channel on $m$ qubits $\cA_\gamma^{\otimes m}$, which models $m$ spontaneous decays on $m$ qubits; each spontaneous decay occurs independently with probability $\gamma$, and each Kraus effect of $\cA_\gamma^{\otimes m}$ has the form 
\[
\tK = \tK_1 \otimes \dots \otimes \tK_m,
\]
where each $\tK_i$ is either $\tA_0$ or $\tA_1$ as defined in Eq.~(\ref{eq:AD-definition}). 
We define ${\rm supp}(\tK)$, the {\em support} of $\tK$, to be the set of all indices $i$ where $\tK_i = \tA_1$, and $\wt(\tK)$, the {\em weight} of $\tK$, to be the cardinality of its support. 
In this section, let our truncated Kraus set be the set of all Kraus effects with weights at most $t$, given by 
\begin{align}
  \Omega:= \{ \tK \in \fK_{\cA_\gamma^{\otimes m}}: \wt(\tK) \le t \} \subset \fK_{\cA_\gamma^{\otimes m}}.  \label{eq:omega-def}
\end{align}
The following lemma gives a lower bound for the trace of our $\tG$-matrix.
\begin{lemma}
\begin{align}
\tr \tG \ge 1 - \bi{m}{t+1}\gamma^{t+1} . \notag
\end{align}
\end{lemma}
\begin{proof}
Applying the definition of $g_{\tA, \tA}$ and exchanging the order of summation, we get 
\[\tr \tG = \sum_{\tA \in \Omega} g_{\tA, \tA}  
  = \sum_{|\alpha\> \in \cB}
\frac{1}{|\cB|}\<\alpha| \sum_{\tA \in \Omega}  \tA ^\dagger \tA |\alpha\>.\]
Since 
\[ \<\alpha| \sum_{\tA \in \Omega}  \tA ^\dagger \tA |\alpha\>
\ge \lambda_{\min} ( \sum_{\tA \in \Omega}  \tA ^\dagger \tA ),\]
hence 
\[ \tr \tG   \ge \lambda_{\min} ( \sum_{\tA \in \Omega}  \tA ^\dagger \tA ) 
  =  \sum_{k =  0}^t \bi{m}{k}\gamma^k (1-\gamma)^{m-k}.\]
Thus for $\gamma \ge 0$, the inequality
\[\tr \tG \ge 1 -   \sum_{k =  t+1}^\infty \bi{m}{k}\gamma^k, \] 
and Taylor's theorem with remainder on $(1+\gamma)^{m}$ gives 
$\tr \tG \ge 1 -   \bi{m}{t+1}\gamma^{t+1} $.
\end{proof}
The inner product 
$\< +_L |  \tA ^\dagger \tB  |  -_L\>$ plays a central role in our analysis of the properties of our deviation matrices $\tG_{\tA,\tB}$ for all $\tA, \tB \in \Omega$. 
Since the gap $g$ is strictly greater than the weight of any Kraus effect from $\Omega$, we have
\begin{align}
\< +_L |  \tA ^\dagger \tB  |  -_L\>
= 
2^{-n} \sum_{\ell = 0 }^n (-1)^\ell \bi{n}{\ell} 
\< D^m_{g \ell }  |  \tA ^\dagger \tB  | D^m_{g \ell } \>.
\label{eq:pm-dicke-decomposition}
\end{align}
Indeed, the absolute value of the inner product $\< +_L |  \tA ^\dagger \tB  |  -_L\>$ is equal to the spectral radius of the deviation matrix $\tG_{\tA,\tB}$ for all $\tA,\tB \in \Omega$.
\begin{lemma}\label{lem:AD-deviation-matrices}
For all $\tA, \tB \in \Omega$, 
\begin{align}
\tG_{\tA,\tB} = \< +_L |  \tA ^\dagger \tB  |  -_L\>
\left( |0_L\>\<0_L| - | 1_L\>\<  1_L |  \right) \notag.
\end{align}
\end{lemma}
\begin{proof}
Since the code has gap $g=t+1$, for all
distinct $j$ and $k$ in $\{0, 1 \}$, the states 
$|j_L\>$ and $|k_L\>$ are supported on Dicke states with excitations spaced $t+1$ apart, it follows that
\[\<j_L | \tA ^\dagger \tB|k_L \>  = 0.\] 
Hence the off-diagonal entries of $\tG_{\tA,\tB}$, given by 
$\< 0_L| \tG_{\tA,\tB}| 1_L\> $ 
and
$\< 1_L| \tG_{\tA,\tB}| 0_L\> $, 
are zero.
The sufficiently large gap $g$ of our code also gives the equalities
\begin{align}
\< 0_L | \tG_{ \tA ,\tB } |  0_L\> &= 
\< 0_L | \tA ^\dagger \tB|  0_L\>  - 
g_{\tA, \tB} \notag\\
&=
\frac{ \< 0_L | \tA ^\dagger \tB|  0_L\>  - 
\< 1_L | \tA ^\dagger \tB|  1_L\>  }{2} \notag\\
&=
\<+_L| \tA ^\dagger \tB |-_L\>, \notag
\end{align}
and
\begin{align}
\< 1_L | \tG_{ \tA ,\tB } |  1_L\> &= 
\< 1_L | \tA ^\dagger \tB|  1_L\>  - 
g_{\tA, \tB} \notag\\
&=
\frac{ \< 1_L | \tA ^\dagger \tB|  1_L\>  - 
\< 0_L | \tA ^\dagger \tB|  0_L\>  }{2} \notag\\
&=
-\<+_L| \tA ^\dagger \tB |-_L\>. \notag
\end{align}
\end{proof}
To complete analyzing our deviation matrices $\tG_{\tA,\tB}$, evaluating the inner products  $\<D^m_{g \ell} | \tA ^\dagger \tB|D^m_{g \ell} \> $ for $0 \le \ell \le n$ is essential. We evaluate these inner products by counting set cardinalities.
\begin{lemma}
\label{lem:inner-product-series-representations}
Let $w$ be a non-negative integer no greater than $m$,
and define the inequality
\begin{align}
 \wt(\tA) \le  w   \le  m -  |\supp (\tA ) \cup \supp (\tB) | + \wt(\tA) .
\label{eq:w-constraint}
\end{align}
For all $\tA,\tB \in \Omega$, 
 $\<D^m_{w} | \tA ^\dagger \tB|D^m_{w} \> $ is 
\begin{align}
\gamma^{\wt(\tA)} 	(1-\gamma)^{w- \wt(\tA) }
    \frac{\bi{  m -  |\supp (\tA ) \cup \supp (\tB) | }{w - \wt(\tA) }}{\bi{m}{w}} 
\delta_{\wt(\tA) , \wt(\tB)}
\label{eq:AD-Dicke-inner-product} .
\end{align}
if (\ref{eq:w-constraint}) holds, and 
 $\<D^m_{w} | \tA ^\dagger \tB|D^m_{w} \> =0$ otherwise.
\end{lemma}
\begin{proof}
For non-negative integers $x$, 
\begin{align}
B_x(\tA) := \{ \bx \in \{0 , 1 \}^m : x_{i } = 0 \ \forall i \in \supp(\tA), \wt(\bx   ) =  x \}. \notag
\end{align}
Now let $x = w - \wt(\tA)$ and $y = w - \wt(\tB)$. 
Then
\begin{subequations}
\begin{align*}
\tA| D^m_w \> &=(1-\gamma)^{x/2} \gamma ^{\wt(\tA)/2} \bi{m}{w}^{-1/2} \sum_{ {\bf x} \in B_{x}(\tA)} | {\bf x } \> , \\
\tB| D^m_w \> &=(1-\gamma)^{y/2} \gamma ^{\wt(\tB)/2} \bi{m}{w}^{-1/2} \sum_{ {\bf y} \in B_{y}(\tB)} | {\bf y } \>.
\end{align*}
\end{subequations}
It follows that $\<D^m_{w} | \tA ^\dagger \tB|D^m_{w} \> $ is equal to 
\begin{align}
&\sqrt{ (1-\gamma)^{x+y}\gamma^{\wt(\tA) + \wt(\tB)} }\bi{m}{w}^{-1} 
\sum_{ 
	\substack{
		{\bf x} \in B_{x}(\tA) \\
		{\bf y} \in B_{y}(\tB) \\
	}
} 
\< {\bf x }  | {\bf y } \>　\notag\\
=&
\bi{m}{w}^{-1} 
\left|	B_{x}(\tA)  \cap B_{y}(\tB) \right|. \label{eq:proof-set-intersection}
\end{align}
The expression in (\ref{eq:proof-set-intersection}) is zero when $B_{x}(\tA)  \cap B_{y}(\tB) = \emptyset$, which happens when any one of the following is true:
\begin{enumerate}
\item
Case $\wt(\tA) \neq \wt(\tB)$: Then $x\neq y$. 
\item 
Case $w < \wt(\tA)$: Then $x < 0$, and $B_{x}(\tA) = \emptyset$. 
\item 
Case $ x > m - |\supp(\tA) \cup \supp(\tB)|$:  
All vectors from the set $B_{x}(\tA)  \cap B_{y}(\tB)$ are necessarily zero on $|\supp(\tA) \cup \supp(\tB)|$ indices. Hence vectors from $B_{x}(\tA)  \cap B_{y}(\tB)$ have a weight of at most 
$m- |\supp(\tA) \cup \supp(\tB)|$. But these vectors must also have a weight of $x$ -- an impossibility.
\end{enumerate}
When the set $B_{x}(\tA)  \cap B_{y}(\tB)$ is non-empty, 
its cardinality is $\bi{m- |\supp(\tA) \cup \supp(\tB)|}{x}$, from which the result follows.
\end{proof}
The inner product $\< + _L  |  \tA ^\dagger \tB| -_L\> $ admits
a Taylor series expansion with respect to the noise parameter $\gamma$, from which we can obtain an upper bound on the total deviation $\epsilon$.
The relevant constants are 
\begin{subequations}
\begin{align}
K_{\tA, \tB} &:= 
\sum_{k \ge n-t} \left| [\gamma^k  ] \< +_L |  \tA ^\dagger \tB  |  -_L\>\right| \gamma_1^{k- (n-t)}, 
\label{eq:K-AB}\\
K &:= \max_ {\tA, \tB \in \Omega  } K_{\tA, \tB}    . \label{eq:K}
\end{align}
\end{subequations}
where $\gamma_1$ is some real number in the open unit interval. 
Indeed, the $\epsilon \le  K \gamma^{n-t}$ if the coefficients 
$[\gamma^k] \< +_L |  \tA ^\dagger \tB  |  -_L\>$ are zero for all $0 \le k \le n -(t+1)$. This is the content of Lemma \ref{lem:eps}, and in its proof
we represent the Taylor series expansions with respect of $\gamma$ of the Dicke inner products 
$\<D^m_{g \ell}  | \tA ^\dagger \tB | D^m_{g \ell}   \>$ 
using polynomials in $\ell$.
Motivated by Lemma \ref{lem:inner-product-series-representations-taylor},
for non-negative integers $a,c,k$ and $\ell$ we define the polynomials with respect to $\ell$ as
\begin{align}
h_{k,a,c} (\ell ) : =  
\frac{ k_{(a) } } { m_{(a+c)}   } (m-\ell g)_{(c)} \bi { \ell g }  { k }.
\label{eq:h-polynomial}
\end{align}
These polynomials are defined so that we have 
\begin{align}
\<D^m_{g \ell}  | \tA ^\dagger \tB | D^m_{g \ell}   \> = 
\sum_{k =0}^{g \ell }h_{k,a,c} (\ell )  \gamma^k
\label{eq:dicke-taylor-series},
\end{align}
where
\begin{align}
c &:=  |\supp(\tA) \cup \supp(\tB)  | - \wt(\tA), \label{eq:c-s-def}
\end{align}
$\tA$ and $\tB$ have equal weights that are positive, and $\ell$ is also positive.
\begin{lemma} 
Let $ \gamma_1$ be a real number in the open unit interval $(0,1)$.
Then for all non-negative reals $\gamma$ no greater than $\gamma_1$, 
\[  
\epsilon \le \max_{ \tA, \tB \in \Omega} |\< +_L |  \tA ^\dagger \tB  |  -_L\>|
\le K \gamma^{n-t}, \]
where $K$ is given by (\ref{eq:K}). 
\label{lem:eps}
\end{lemma}
\begin{proof}
The first inequality of this lemma follows directly from Lemma \ref{lem:AD-deviation-matrices}. 
First note the inner product 
$\< +_L |  \tA ^\dagger \tB  |  -_L\> $ is zero when (i) $\wt(\tA) = \wt(\tB) =0$, and when (ii) $\wt( \tA) \neq \wt(\tB)$. Hence we focus on Kraus effects $\tA$ and $\tB$ of equal weight $a$ where $1 \le a \le t$.

Using Lemma \ref{lem:inner-product-series-representations-taylor} and Lemma \ref{lem:inner-product-series-representations} on the decomposition given by Eq.~(\ref{eq:pm-dicke-decomposition}), for non-negative integers $k$, the Taylor series (\ref{eq:dicke-taylor-series}) holds. 

Since these Taylor series are finite, we have 
that $\< +_L |  \tA ^\dagger \tB  |  -_L\>$ is equal to
\begin{align}
&2^{-n} \sum_{\ell=1}^{n}	\bi{n}{\ell} (-1)^{\ell}
\<D^m_{g \ell}|  \tA^ \dagger \tB | D^m_{g \ell} \> \notag\\
=&
2^{-n}\sum_{k \ge 0} 
\left(
	 \sum_{\ell=1}^{n}	\bi{n}{\ell}  (-1)^{\ell} (-1)^{k-a}  h_{k,a,c} ( \ell ) \gamma^k
\right)  \label{eq:final-poly-form} ,
\end{align}
where $c$ is as given by (\ref{eq:c-s-def}).
In Eq.~(\ref{eq:final-poly-form}), we interchange the order of the summations, which is valid because the Taylor series (\ref{eq:dicke-taylor-series}) is a finite sum. 
Hence $[\gamma]^k\< +_L |  \tA ^\dagger \tB  |  -_L\>$ is equal to
\begin{align}
2^{-n}  \sum_{\ell=1}^{n}	\bi{n}{\ell} (-1)^{\ell}
(-1)^{k-a}  h_{k,a,c} ( \ell ) \label{eq:area-47}.
\end{align}
Now the polynomials $h_{k,a,c}$ satisfy the equality $h_{k,a,c}(0) = 0$ for all non-negative integers $k,a,c$.
Hence the expression (\ref{eq:area-47}) is equivalent to the expression
\begin{align}
2^{-n} (-1)^{k-a}  \sum_{\ell=0}^{n}	\bi{n}{\ell} (-1)^{\ell}
 h_{k,a,c} ( \ell ) \label{eq:area-48}.
\end{align}
Since the polynomials $h_{k,a,c}$ are of order $k+c$ with respect to the parameter $\ell$ and $c \le t$, their maximum order is $n-1$ for all $0 \le k \le n-(t+1)$.
Lemma \ref{lem:cute-binomial-identity} then implies that all the bracketed terms in the right hand side of Eq.~(\ref{eq:final-poly-form}) are zero  when $k \le n-(t+1)$. 
\end{proof}
Let us denote $\<j_L| \tA ^\dagger \tA | j_L\>$ explicitly as a function of the noise parameter $\gamma$ using the function $f_{j,\tA} (\gamma)$. 
From this function's decomposition as given by Lemma \ref{lem:inner-product-series-representations},
it is clear that for non-negative $ \gamma \le \frac{1}{2}$ and $t \le \frac{m}{2}$, the function $f_{j,\tA}$ is monotone increasing with respect to $\gamma$. Hence for $\gamma$ smaller than $\min \{ \gamma_1, \frac{1}{2} \}$, 
the minimum eigenvalue of our $\tG$-matrix is $\frac{L}{2} \gamma^{t}$, where $L$ is defined as 
\begin{align}
L := \min_{\substack{ \tA \in \Omega \\ j\in \{0,1\} } }
     f_{j, \tA} ( \gamma_1 ) . \label{eq:L-def}
\end{align}
This is the content of the following lemma.
\begin{lemma}  \label{lem:lmin} 
Let $\gamma_1$ be a real number no greater than $\frac{1}{2}$, and let $L$ be as defined in (\ref{eq:L-def}).
Let 
\begin{align}
\gamma_0 = 
m^{-t/(n-2t)} \left(
\frac{L}{2K}
\right)^{1/(n-2t)}
, \notag
\end{align}
and suppose that $\gamma_0 \le \gamma_1$. 
Then for all non-negative reals no greater than $\gamma_0$, 
we have the lower bound
\[\lambda_{\min} (\tG) \ge \frac{L}{2} \gamma^t.\]
\end{lemma}
\begin{proof}
Clearly $\displaystyle \min_{\tA \in \Omega} g_{\tA,\tA} \ge L \gamma ^t$.
Each row in $\tG$ has at most $\bi{m}{t}$ non-zero entries, each entry with magnitude at most $K \gamma^{n-t}$. 
Since $\gamma < \gamma_0$, the Ger\v sgorin circle theorem implies that 
\[ 
\lambda_{\min}(\tG) \ge L \gamma^t - \bi{m}{t} K \gamma^t  \gamma^{n-2t}.
\]
Since $\gamma_0$ is the minimum of the set $\{ \gamma_0 , \gamma_1, \frac{1}{2} \}$, it follows that 
\[ 
\lambda_{\min}(\tG) \ge L \gamma^t - m^t K \gamma^t  
\left( \frac{L}{2K} m^{-t} \right)
= \frac{L}{2} \gamma^t.
\]

\end{proof}
Piecing our results together, we can quantify the error 
after using our gnu codes to correct spontaneous decay errors:
\begin{theorem}\label{thm:AD}
Suppose that the non-negative reals $\gamma_0$ and $\gamma_1$ are such that the assumption of Lemma \ref{lem:lmin} holds. Let $K$ and $L$ be given by 
(\ref{eq:K}) and (\ref{eq:L-def}) respectively. Then for all $\gamma \le \gamma_0$, 
a gnu code with gap $g=t+1$, occupancy number $n > 3t$, scaling factor $u \ge 1 + \frac{t}{gn}$, and a length $m=gnu$ has a worst case error with respect to the noisy channel 
$\mathcal A_\gamma ^{\otimes m}$ at most
\begin{align}
1 - 
	\frac{1-\bi{m}{t+1} \gamma^{t+1} - |\Omega|^2 K \gamma^{n-t}     }
		{1+
\frac{ 2(|\Omega |-1) |\Omega|^2 K }{  L  }
\gamma^{n-2t}}
\label{eq:main-result}
\end{align}
\end{theorem}
\begin{proof}
From the above we have the upper bound $\epsilon \le K \gamma^{ n-t}$, and the lower bounds $\lambda_{\min} (\tG) \ge \frac{L}{2} \gamma^t$ and $\tr(\tG) \ge 1- \bi{m}{t+1} \gamma^{t+1}$.
Using Theorem \ref{thm:pkl}, and by choosing the rescaling factor 
\[
\eta =  \frac{ (|\Omega |-1) |\Omega|^2 K \gamma^{n-t}   }
{ \frac{ L} {2} \gamma^t }  = 
\frac{ 2(|\Omega |-1) |\Omega|^2 K }
{  L  } \gamma^{n-2t}   ,
\]
we get the result.
\end{proof}
Note that the upper bound on the error in the above theorem converges to zero at the appropriate rate in the limit as $\gamma$ approaches zero. Thus Eq.~(\ref{eq:PI-code}) gives a family of $t$-AD PI codes. 

\begin{figure}[ht]
\begin{minipage}[b]{\linewidth}
\centering
\includegraphics[width=\textwidth]{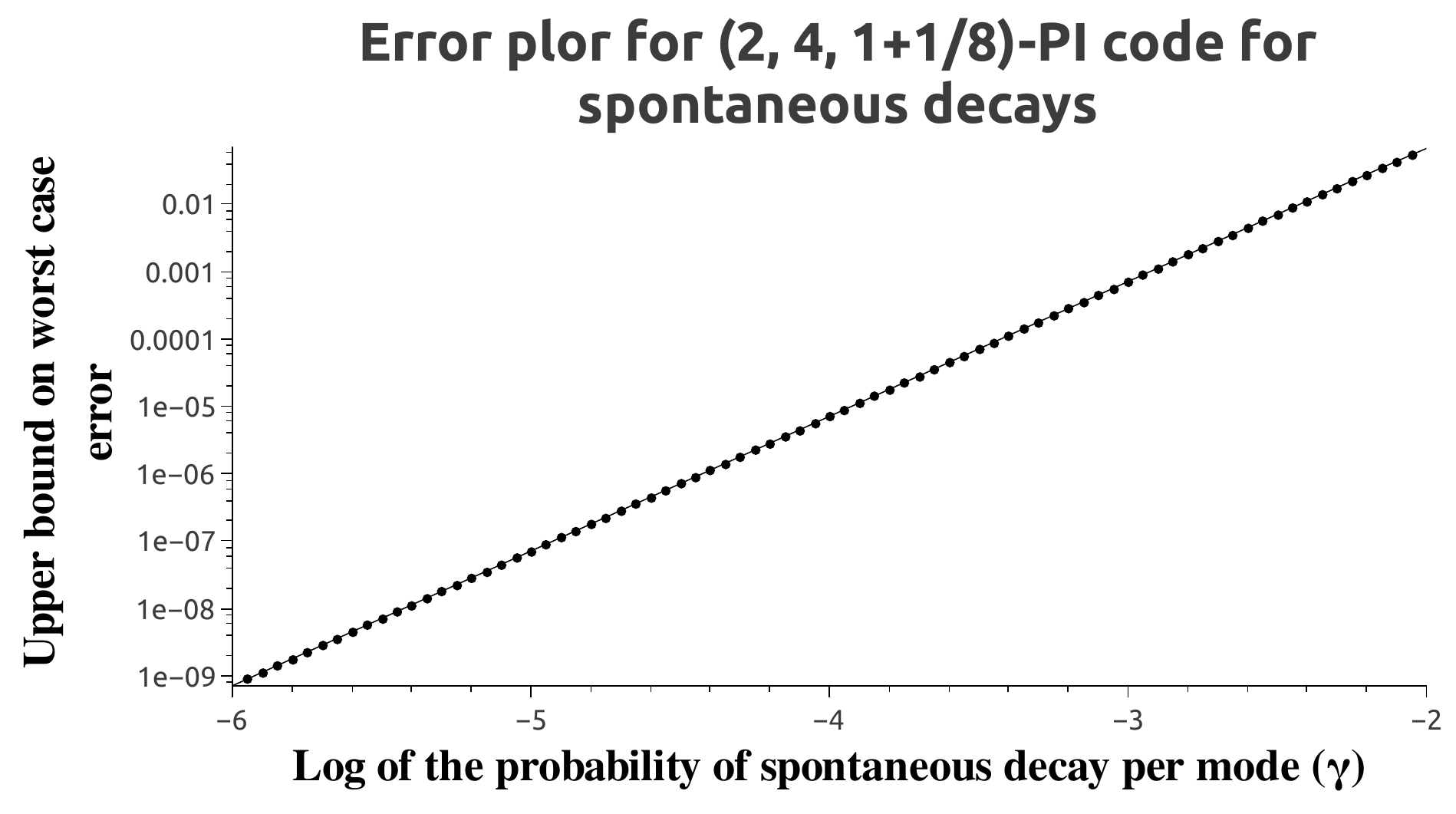}
\end{minipage}
\begin{minipage}[b]{\linewidth}
\centering
\includegraphics[width=\textwidth]{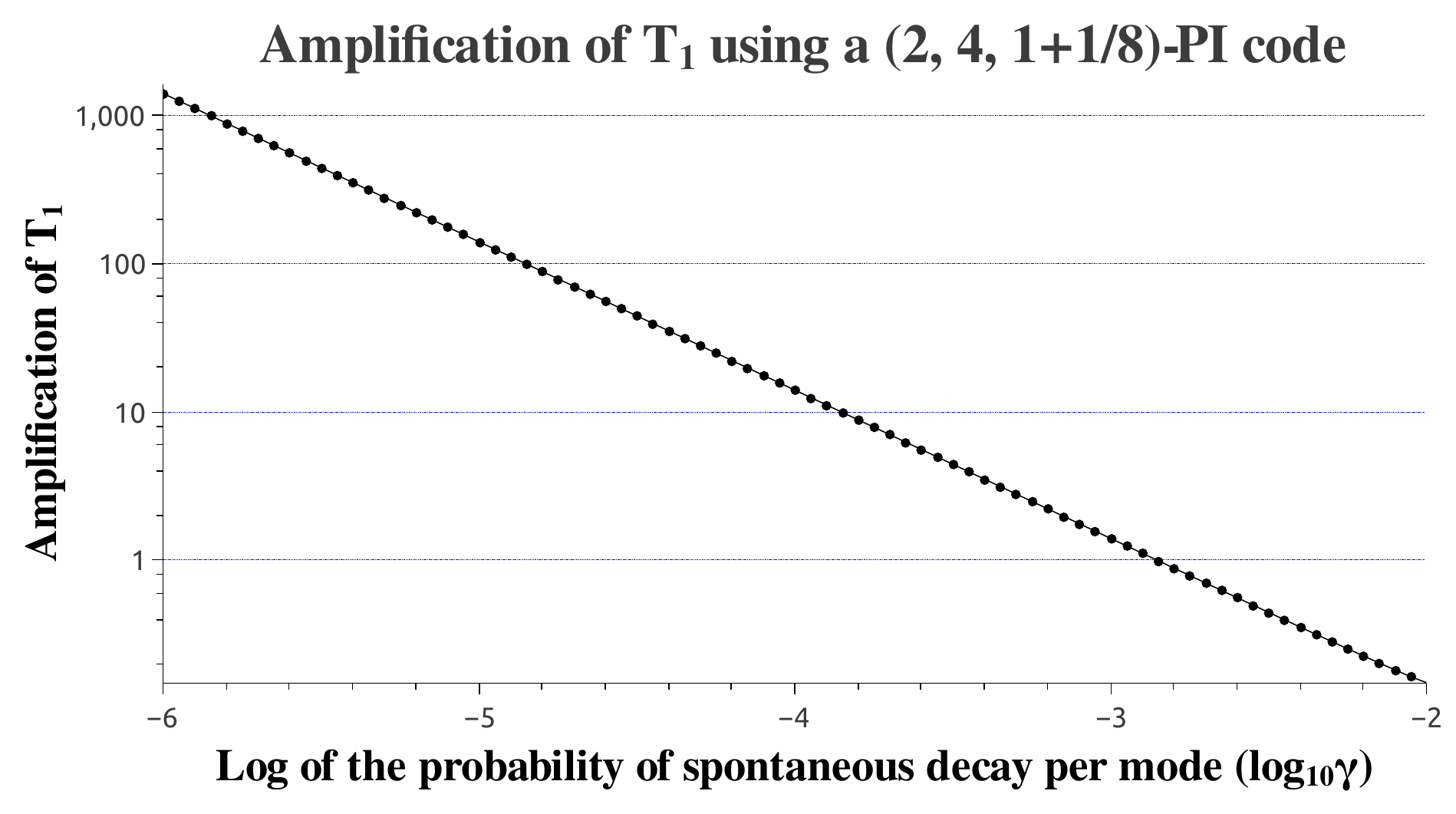}
\end{minipage}
\caption{The performance of a $(2,4,1+1/8)$-PI code on 9 qubits is quantified, both in terms of upper bounds of the worst case error
({\bf Top}) and lower bounds on amplification of the effective $T_1$ 
({\bf Bottom}). 
\label{fig:1}
}
\end{figure}

We depict the performance of our $(2,4,1+1/8)$-PI code on 9 qubits in terms of error rates and amplifications of the of effective $T_1$ times in Fig.~\ref{fig:1}.

\section{Discussions}
\label{sec:discussions}
 In this \paper, we construct $t$-AD codes and codes correcting arbitrary $t$-sparse errors from our gnu code family. 
Since gnu codes lie within the ground state of ferromagnetic Heisenberg models without an external magnetic field,
one might hope that such codes are viable candidates in realizing a viable quantum memory \cite{OuF14}. 

To determine whether gnu codes are suitable for practical use, many difficulties remain to be overcome. 
Easily implementable encoding and decoding protocols for these gnu codes have to be devised, and the explicit quantum circuits for the error correction procedure also remain to be determined. 
The possibility of having fault-tolerant PI codes remains to be investigated, because the standard framework of fault-tolerant quantum computing \cite{AGP05} does not apply directly to non-stabilizer codes. 
Also the relationship between our work and other results on implementation of permutation-invariant quantum circuits \cite{HSa11,MHT12,KBS13} remains to be more thoroughly investigated.
We leave these challenges amongst many others for future study. \\

\section{Acknowledgments}
Y.~Ouyang is especially grateful for discussions with Joseph Fitzsimons, Tommaso Demarie and Jon Tyson, and comments from Debbie Leung and Ashwin Nayak. The author acknowledges support from the Ministry of Education, Singapore.

\bibliography{../../../mybib}{}
\addcontentsline{toc}{\paper}{Bibliography}
\bibliographystyle{ieeetr}
\end{document}